# On the Generation of Phononic Frequency Combs Using Defect Modes of Phononic Crystals

Suhas Suresh Bharadwaj
*Department of Electrical
and Electronics Engineering*
*BITS Pilani, Dubai Campus*
Dubai, United Arab Emirates
f20230029@dubai.bits-pilani.ac.in

Murtaza Rangwala
*Department of Electrical
and Electronics Engineering*
*BITS Pilani, Dubai Campus*
Dubai, United Arab Emirates
f20240520@dubai.bits-pilani.ac.in

Adarsh Ganesan
*Department of Electrical
and Electronics Engineering*
*BITS Pilani, Dubai Campus*
Dubai, United Arab Emirates
adarsh@dubai.bits-pilani.ac.in

*Abstract*—This paper proposes a method for generating phononic frequency combs (PFCs) using defect-localized modes in a two-dimensional hexagonal phononic crystal. Localized vibration modes from a singular point defect produce evenly spaced spectral lines corresponding to PFCs. Numerical modelling reveals robust energy transfer under a single-tone drive, generating spectral sidebands. These results demonstrate defect engineering in phononic crystals as a tunable platform for PFC generation with significant applications in high-resolution sensing, timing, and quantum-acoustic technologies.

## I. Introduction

Frequency combs are spectral structures composed of a series of discrete, equally spaced frequency lines and have been increasingly used by the optics community as metrological and spectroscopic tools [1]. In 2017, Ganesan et al. [2] discovered analogous mathematical structures in the phononic domain named phononic frequency combs (PFC). The generation mechanisms of PFCs are fundamentally rooted in nonlinear modal interactions, resulting from parametric resonance between coupled mechanical modes [2]–[4]. Besides nonlinear modal coupling, follow-up experimental studies have explored the generation of PFCs via optomechanical interactions [5]–[8].

In this paper, we theorize the possibility of generating PFCs in phononic crystals. Phononic crystals are artificially engineered materials with periodic variations in elastic properties designed to control mechanical wave propagation [9], [10]. The bandgaps that arise within these phononic crystals are termed as phononic bandgaps. These bandgaps are the result of destructive interference of mechanical waves suppressing vibrations and thermal phonons in specific frequency ranges. Introducing intentional defects within such periodic lattices locally breaks translational symmetry, giving rise to defect modes whose vibrational energy is spatially confined near the defect region and whose frequencies lie within the bandgap [11], [12]. Previous works have largely studied phononic crystals with defects in the context of energy localization [13], [14] and engineering high-quality factor mechanical modes [15], [16]. However, no prior work has examined the generation of PFC arising from defect modes embedded within the bandgaps of phononic crystals. In this paper, we explore the possibility of harnessing these defect-localized modes as a platform for the generation of PFCs, using a two-dimensional hexagonal phononic crystal lattice.

In this work, we design and analyze a two-dimensional hexagonal phononic crystal lattice in COMSOL Multiphysics to investigate the formation of localized defect modes within the phononic bandgap. By introducing a central point defect and performing eigenfrequency analyses, we identify defect-confined vibrational modes and extract their corresponding frequencies. These modal parameters and frequencies are then used in a Python-based post-processing model to simulate and analyze PFC generation. This study successfully demonstrates that defect modes can act as localized oscillators, providing a potential pathway for tunable PFC generation.

## II. Simulation Setup

To study PFC generation using defect modes of phononic crystals, we first study the phononic bandgaps of the crystal lattice. Therefore, we introduce defects in the lattice and observe the emergence of defect modes within the bandgaps. By appropriately choosing the lattice geometrical parameters and defect size, we aligned these defect modes in the frequency ratio of 1:2, and therefore making them suitable for PFC generation. This section details the simulation setup for these analyses as follows.

### A. Construction of phononic crystal unit cell and lattice

We construct a two-dimensional hexagonal phononic crystal lattice in COMSOL Multiphysics, beginning with a regular hexagon of side length $a$ = 80 $\mu$m (Fig. 1(a)). Circular sectors of radius $R$ are removed from each vertex to form the modified hexagonal element shown in Fig. 1(b). Tethers of length $l/2$ are then attached to each face of the hexagon, producing a star-like structural motif (Fig. 1(c–d)). By repeating this basic element periodically, a two-dimensional hexagonal lattice pattern is generated (Fig. 1(e)), where the dashed rectangle indicates the primitive unit cell. The extracted unit cell (Fig. 1(f)) exhibits a geometry defined solely by the parameters $R$ and $l$ since both are by definition a function of $a$. To model a finite sample, the unit cell is periodically replicated (Fig. 1(g)). By varying the tether length $l$ and the circular sector radius $R$, multiple geometric configurations of the same lattice can be achieved with only minor variations in the overall dimensions. For the purposes of our simulation, we constructed a lattice with a unit cell having $R = 39$ μm, and tether length of 16 μm (Fig. 1(g-h)). While the material of unit cell is defined as Si (Silicon), the gaps are filled with air.

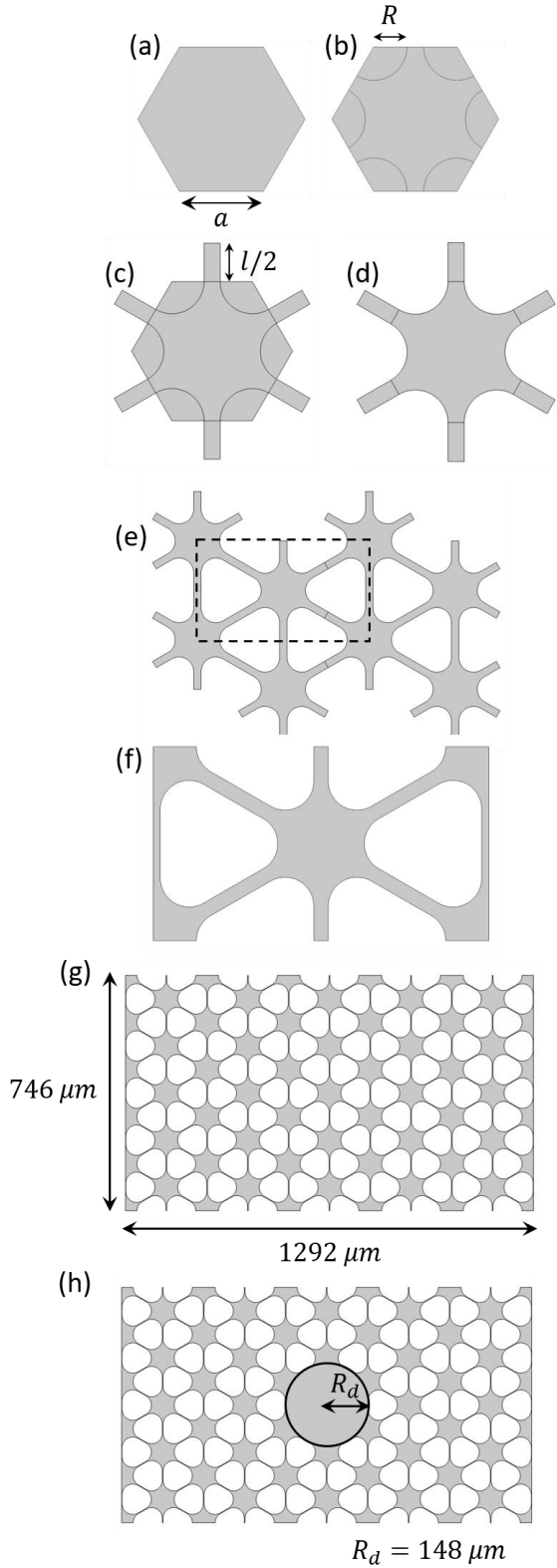

Fig. 1. Construction of the unit cell and lattice of the phononic crystal. (a) A regular hexagon of side length a; (b) Circular sectors of radius R are centered at each vertex of the hexagon; (c) Tethers of length l/2 are attached to the hexagon; (d) Removal of the sectors to obtain a geometry element; (e) Tiling the elementary patterns to produce a lattice; (f) Unit cell of arbitrary $R$ and tether length value which is then varied to obtain the desired geometry; (g) Final phononic crystal lattice geometry used for the present study with a unit cell of $R = 39$ μm and tether length of 16 μm; (h) Phononic crystal lattice with circular defect of radius $R_d$ (Here, $R_d = 148$ μm).

### B. Defect engineering

To the constructed lattice in Fig. 1(h), we introduced a circular point defect at the center of the lattice with a radius of 148 μm (Fig. 1(i)). This point defect radius is chosen so as to completely fill the air gap holes surrounding the central hexagon of the lattice and is made up of Si (Silicon).

### C. Bandgap and Eigenfrequency simulation

Phononic wave propagation was modeled using the structural mechanics module in COMSOL Multiphysics. For each geometry we performed eigenfrequency studies using the ARPACK eigenvalue solver to obtain the lowest 600 eigenmodes. These eigenfrequencies were used to construct phononic band diagrams, and representative mode shapes for individual eigenfrequencies were plotted to study the vibrational modes in the lattice. The hexagonal lattice (Fig. 1(h)) was designed in such a way that the central frequencies of the two prominent bandgaps exhibit a 1:2 ratio. To investigate defect-induced states, the eigenfrequency analysis was repeated for a lattice containing a central circular defect with radius $R_d = 148$ μm.

### D. Mathematical model and simulation of PFC

$$\ddot{x}_1 + 2\gamma_1 \dot{x}_1 + \omega_1^2 x_1 + \alpha_{22} x_2^2 = F\cos(\omega_D t), \quad (1)$$

$$\ddot{x}_2 + 2\gamma_2 \dot{x}_2 + \omega_2^2 x_2 + \alpha_{12} x_1 x_2 = 0. \quad (2)$$

To simulate the dynamics leading to PFC formation, the nonlinear coupled modal equations (Eqs. (1)–(2)) were solved numerically using Python's `solve_ivp()` function within the *SciPy* library. These equations describe a two-degree-of-freedom nonlinear oscillator model, where $x_1(t)$ and $x_2(t)$ denote the modal displacements of the dominant defects (mode 1 and mode 2). The parameters $\omega_{1,2}$ represent the natural angular frequencies of these modes, while $\gamma_{1,2}$ denote the corresponding damping coefficients derived from the quality factors $Q_{1,2}$ obtained through eigenfrequency analysis in COMSOL Multiphysics. The external single-tone drive applied to mode 1 is described by $F\cos(\omega_D t)$. The nonlinear terms $\alpha_{22} x_2^2$ and $\alpha_{12} x_1 x_2$ represent quadratic and intermodal coupling effects arising from geometric and material nonlinearities [17], [18]. The integration was performed over $10^{-4}$ s using an adaptive Runge–Kutta method.

The primary mode frequency was set to $\omega_1 = 31.295$ MHz and the secondary mode to $\omega_2 = 16.318$ MHz, yielding a ratio $\omega_2/(\omega_1/2) = 1.043$, which satisfies the near-resonant condition necessary for efficient three-wave mixing and PFC formation [17]. The quality factors were chosen as $Q_1 = 1000$ for the driven mode and $Q_2 = 10^6$ for the secondary mode, ensuring high energy dissipation in the primary mode and very low energy dissipation in the secondary mode to initiate intermodal energy exchange and long-lived comb dynamics. The drive frequency was set at $\omega_D = 31.296$ MHz, closely matching the center of the existence band ($\omega_D^{center} = 31.630$ MHz), with the range of comb existence $R = 669.1$ kHz characterizing the bandwidth over which comb generation exists. A critical parameter in the observation of phononic frequency comb formation is the minimum force threshold ($F_{min}$) required to initiate nonlinear intermodal energy transfer. In our simulation, this threshold was analytically calculated based on the system's modal frequencies, damping rates, and nonlinear coupling coefficients, yielding $F_{min} = 0.43$ $mN$. To better verify and observe comb dynamics, the actual drive force applied on a 10 $\mu m$ thick phononic crystal was set to $F = 1.3$ $mN$, approximately three times the calculated threshold. This ensures the system operates well within the range where multi-mode energy exchange and nonlinear spectral broadening become clearly visible. Setting these parameters for the simulation, we performed time-domain analyses for both oscillators and applied Fourier transforms on the steady-state signals to extract the PFC spectra. This simulation approach effectively replicates the intermodal coupling behavior observed experimentally in phononic crystal structures with defect-induced localized modes.

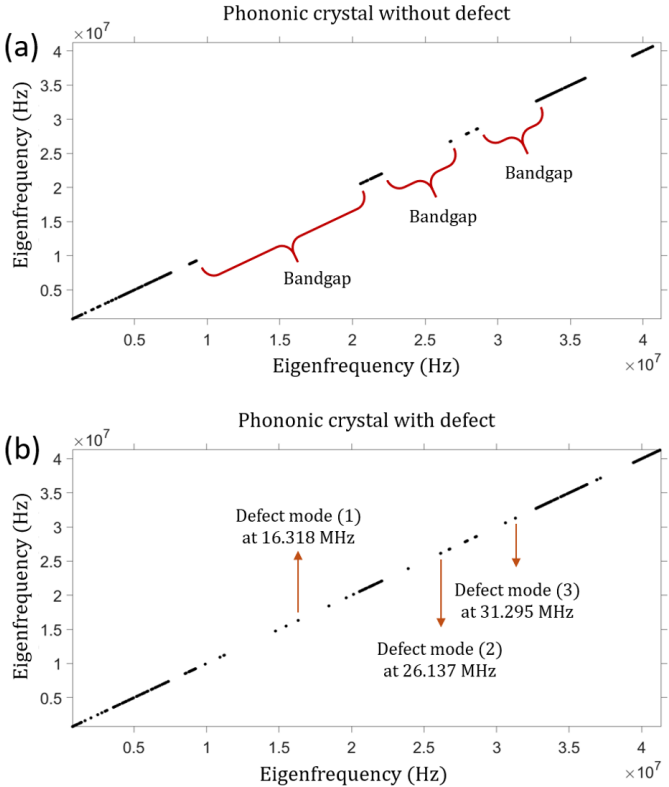

Fig. 2. Plots showing the first 600 eigenmodes of the lattice (a) without defect; (b) with circular defect of radius $R_d = 148$ $\mu$m.

## III. RESULTS AND DISCUSSION

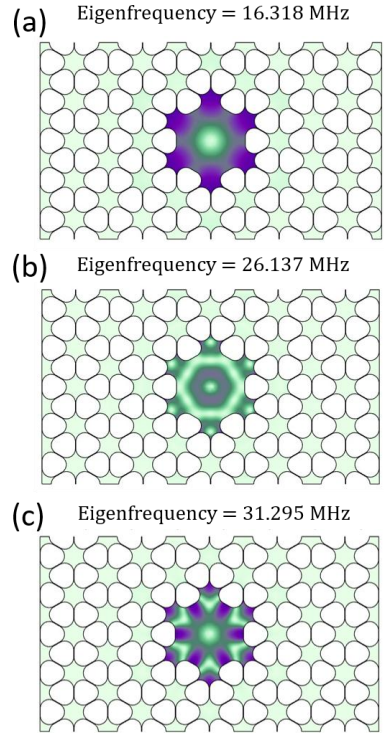

Fig. 3. Deformation plots of the three defect modes identified with eigenfrequencies (a) 16.318 MHz; (b) 26.137 MHz and (c) 31.295 MHz.

Fig. 2(a) shows the bandgaps of the defect-less lattice. These bandgaps arise from destructive interference of elastic waves, which suppress vibrational motion over specific frequency ranges. When an eigenfrequency analysis is performed on the lattice containing the circular defect, additional modes appear inside the bandgap (Fig. 2(b)), where the defect-less lattice exhibits a complete absence of states. Further investigation of the corresponding mode shapes indicates that a few of these defect-induced modes are spatially localized denoting that the deformation and vibrational energy are concentrated within the defect region and decay into the surrounding crystal (Fig 3). Such localized resonances arise solely from the introduced defect. A few of the other eigenmodes that appear within the bandgap do not show any mode localization around the defect and hence are considered as extended modes.

Introducing defects into the lattice breaks the perfect periodicity and creates defect-modes whose frequencies lie within the bandgap and its vibrational energy spatially confined to the defect. Since the waves at these defect mode frequencies cannot propagate through the perfect lattice, the defect acts like a mechanical cavity. It traps energy localized in the defect region and amplifies the motion and reducing losses due to wave propagation [19], [20]. We use these defects within the lattice to act as a cavity for localized resonance that facilitates nonlinear interaction like three-wave mixing ultimately giving rise to PFCs.

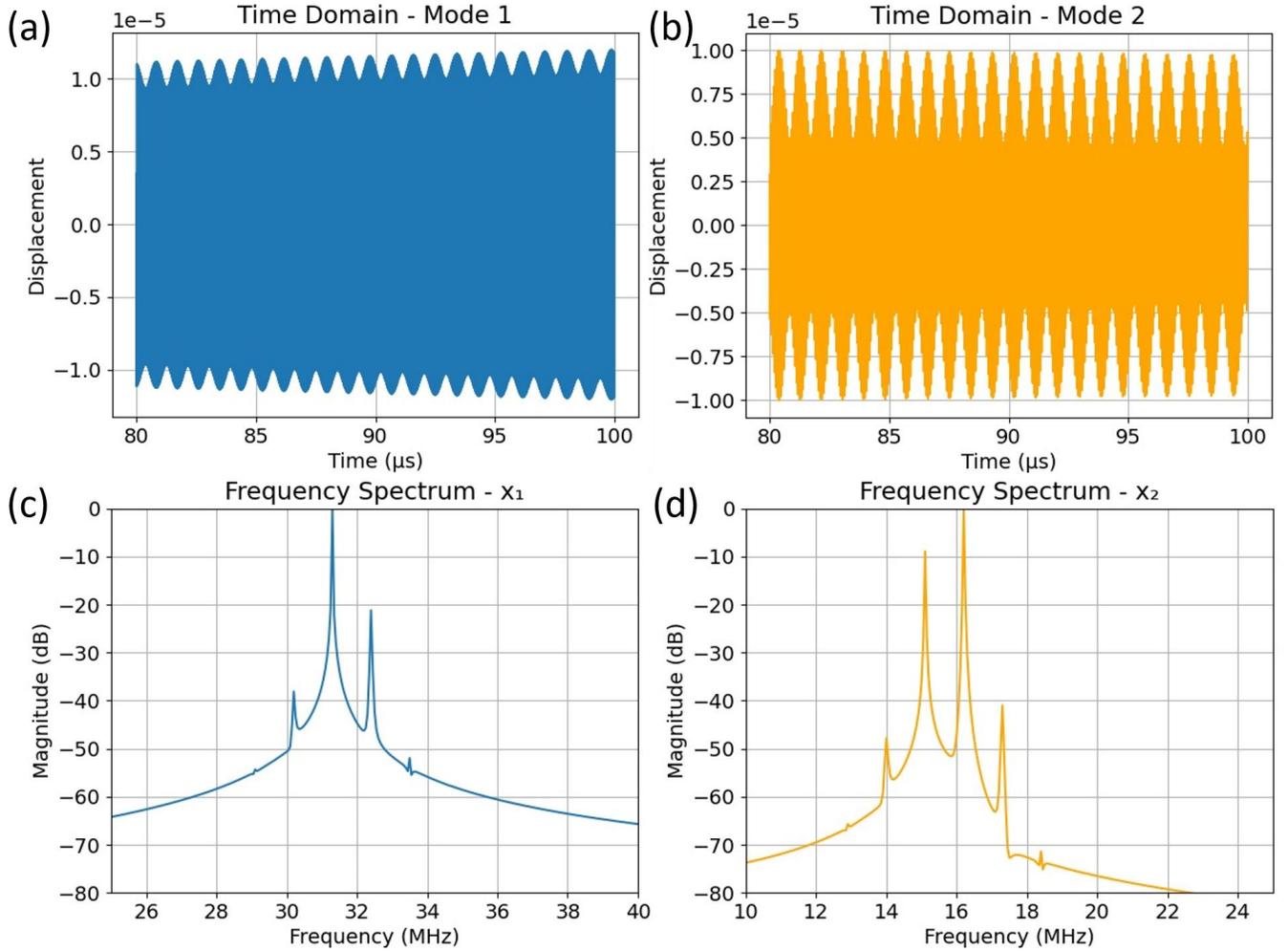

Fig. 4. Numerical simulation results of phononic frequency comb generation through nonlinear modal coupling. (a) Time-domain displacement response for Mode 1 (primary driven mode at 31.295 MHz) showing periodic oscillations with clear amplitude modulation; (b) Time-domain displacement response for Mode 2 (secondary, subharmonic mode at 16.318 MHz); (c) Frequency spectrum of Mode 1 displaying a central peak at the drive frequency with symmetric sidebands characteristic of PFC formation; (d) Frequency spectrum of Mode 2 showing multiple equally spaced spectral lines around the frequency, confirming successful PFC generation through defect-induced intermodal coupling.

The time-domain simulations on the Python simulator with the aforementioned parameters reveal periodic oscillations in both displacement channels (mode 1 and mode 2 depicted in Fig. 4(a-b)). Mode 1 (the primary *driven* mode) and Mode 1 (the secondary *excited* mode) both reach steady-state amplitudes with clear waveform modulation, showing effective energy transfer and coupling between the modes. Fourier transform analysis of these signals (Fig. 4(c–d)) demonstrates the emergence of distinct PFCs. The frequency spectrum of $x_1$ (primary mode) exhibits a central peak at the drive frequency (∼31.3 MHz), accompanied by equally spaced sidebands which is characteristic of PFC formation. Similarly, $x_2$ (secondary mode) shows a rich spectrum around the frequency (∼16.3 MHz), with symmetric sidebands confirming parametric resonance and nonlinear mixing. The visualized comb spacing, symmetry, and spectral line amplitudes are in close agreement with previous simulation and experimental results and confirm the successful numerical emulation of PFC dynamics in the designed structure through the introduction of defects.

## IV. CONCLUSION

This study forms a thorough framework for producing PFC via defect-induced mode localization in two-dimensional hexagonal phononic crystals. We successfully show that strategically engineered defect modes facilitate robust PFC formation by integrating numerical eigenfrequency analysis with a nonlinear coupled oscillator model. The key contributions of this research include: (i) a rigorous methodology for designing bandgap-tunable phononic crystals with controllable defect modes and (ii) the first theoretical demonstration of PFC generation through engineered defect modes in phononic crystal lattices.

The significance of this research extends to emerging technological fields where PFCs offer advantages over optical combs to engineer high-sensitivity tuneable MEMS sensors to act as thermometers, accelerometers and magnetometers. By using defect-mode localization, we achieve a comb source with high spectral purity which can be used for high-resolution sensing, timing references, and other quantum acoustic devices. Future work must therefore focus on further study of PFC generation through multi-defect configurations for enhanced bandwidth, and exploring different lattice defects for optimal PFC generation.